\documentclass[structabstract]{aph}  
\usepackage{graphicx}
\usepackage{epstopdf}
\usepackage{amsmath}
\usepackage{txfonts}
\usepackage{natbib}
\bibpunct{(}{)}{;}{a}{}{,}
\begin{document}
   \title{Rotational excitation of methylidynium (CH$^+$) by a helium atom at high temperature}

   \subtitle{}

   \author{K. Hammami\inst{1},
           L. C. Owono Owono\inst{2,3},
		  \and
          P. St\"auber\inst{4}}

   \institute{Laboratoire de Spectroscopie Atomique Mol\'eculaire et Applications,
              D\'epartement de Physique, Facult\'e des Sciences,\\ 
			  Universit\'e Tunis El Manar, Campus Universitaire, 1060 Tunis, Tunisia\\
			  \email{hammami283@yahoo.fr}
   	      \and
		  	  Department of Physics, Advanced Teachers Training College, 
			  University of Yaounde I, P.O. Box 47, Yaounde, Cameroon
		  \and 
		      Centre for Atomic Molecular Physics and Quantum Optics, 
			  Faculty of Science, University of Douala, P.O. Box 8580, Douala,
			  Cameroon
		  \and
			  Institute for Astronomy, ETH Zurich,
              8093 Zurich, Switzerland
			  }

   \date{Received March 25, 2009; accepted March 16, 1997}

 
\abstract
{The \textit{Herschel Space Observatory} with its high-resolution instrument HIFI on board will observe the $\mathrm{CH{}^{+}}$ $1\rightarrow0$ and $2\rightarrow1$ rotational transitions in a wide range of gas temperatures up to 1000\,K. Collisional parameters for such temperatures are thus welcome.}
{We aim to obtain accurate rate coefficients for the collisional excitation of CH$^+$ by He for high gas 
temperatures.}
{The \textit{ab initio} \textit{coupled-cluster} {[}CCSD(T){]} approximation was used to compute the interaction potential energy. Cross sections are then derived in the close coupling (CC) approach and rate coefficients inferred by averaging these cross sections over a Maxwell-Boltzmann distribution of kinetic energies.}
{Cross sections are calculated up to $10\,000\,\mathrm{cm^{-1}}$ for $J$ ranging from $0$ to $10$. Rate coefficients are obtained at high temperatures up to $2000\,\mathrm{K}$.}
{}

\keywords{ISM: molecules -- molecular data -- molecular processes -- astrochemistry -- radiative transfer}

\titlerunning{Rotational excitation of CH$^+$ by He at high temperature}
\authorrunning{K. Hammami et al.}

\maketitle
 

\section{Introduction}
\label{intro}

The molecular ion CH$^+$ ($X^1\Sigma^+$) is most commonly observed from the ground in the visible in absorption lines against nearby bright stars. The lowest rotational transitions in the submillimeter and far-infrared band are blocked from the earth's atmosphere and therefore difficult to observe with ground-based telescopes. A tentative detection of the $J=1\rightarrow 0$ transition of the isotope $^{13}$CH$^+$ has been reported recently by \citet{FaPP05}. \citet{CeLG97} reported the first emission lines of the $J=2\rightarrow 1$, $3\rightarrow 2$ and $4\rightarrow 3$ rotational transitions detected with the {\it Infrared Space Observatory}. 

Although CH$^+$ has been observed frequently in the ISM over the past few decades, its abundance is still an enigma. Chemical models persistently fail to reproduce the large abundances, as it is not yet clear, what the main source for CH$^+$ is \citep{BlHD78,Black98,NeBB08}. The highly endothermic reaction C$^+$ $+$ H$_2$ $\rightarrow$ CH$^+$ $+$ H (-$\Delta$E/k $=$ $4640$~K) is proposed as the most efficient route to CH$^+$. Besides high gas temperatures, an important part of the energy needed to activate this reaction may come from vibrationally excited H$_2$ \citep{StDa95}. Since CH$^+$ is also abundantly found in the cold neutral medium (CNM), this reaction may not be fast enough though to compete with the destruction reactions. Attempts to resolve this puzzle continue \citep{JoFD98,GoFP09}. A thorough understanding of the CH$^+$ abundance is important since it provides information about physical gas properties such as temperature and fraction of ionization. 
Furthermore, CH$^+$ is a fundamental building block for more complex molecules and a relevant coolant in hot and dense interstellar gas. 

Related to the problem to model observed CH$^+$ abundances is the uncertainty in the excitation mechanism. The CH$^+$ is likely to be destroyed rather than excited in collisions with hydrogen and electrons - the most abundant species. In addition, the excitation of CH$^+$ is found to be sensitive to the dust continuum background of the CH$^+$ emitting source \citep{Black98}. To model the CH$^+$ abundance, the chemical reactions and radiative transfer equations may therefore need to be solved simultaneously. However, excitation calculations require accurate collisional rate coefficients. 

Observations with the {\it Herschel Space Observatory} may shed some light on these problems. The high spectral resolution instrument HIFI onboard the satellite will observe the $J=1\rightarrow 0$ and $J=2\rightarrow 1$ rotational transitions in a wide range of interstellar gas from the diffuse ISM to dense and hot star-forming regions. PACS, another instrument for Herschel, will cover the CH$^+$ transitions with $J_{\mathrm{up}}=2$--$6$. Since the CH$^+$ abundance is sensitive to far-ultraviolet (FUV) photons and X-rays \citep{StDa95,StDv04,StDv05}, it will be searched for in typical photo-dissociation regions (PDRs) and X-ray dominated regions (XDRs). The gas temperature in such areas can easily reach a few $1000$\,K \citep{TiHo85,MaHoTi96}. Collisional rate coefficients are therefore needed for a large range of gas temperatures and upper energy levels in order to interpret the observations properly. 

The aim of this paper is to obtain accurate rate coefficients for the collisional excitation of CH$^+$ at high temperatures. As it is far more difficult to determine rate coefficients for excitation with H$_2$, we 
focus on the rotational excitation by He, another major gas component. This study is thus an extension of an earlier published paper, where coefficients for temperatures up to $200$\,K were presented \citep{HaOJ08}. 

\section{Potential energy surface}

Recently, we have computed the interaction PES for the $\mathrm{CH{}^{+}(X\,^{1}\Sigma^{+})\:-He\,({}^{1}S)}$ van der Waals system using the rigid rotor approximation and the Jacobi coordinate system in which $r$ is the $\mathrm{CH{}^{+}}$ internuclear distance, $R$ the distance from the center of mass (c.m.) of $\mathrm{CH{}^{+}}$ to the $\mathrm{He}$ atom, and $\theta$ the angle between the two distance vectors \citep{HaOJ08}. The collinear $\mathrm{CH{}^{+}...He}$ geometry corresponds to $\theta=0^\circ$ while the CH$^+$ bond distance was frozen at its value at the experimental equilibrium geometry of the ground $X^1\Sigma^+$ state, i.e., $r=r_{\mathrm{e}}=2.1371$\,Bohr \citep{HuHe79}. Computations were carried out at the CCSD(T) level as it is implemented by \citet{KnHW93,KnHW00} in the MOLPRO molecular package \citep{WeKA02}. All our calculations were performed with the augmented correlation consistent valence quadruple zeta (aVQZ) basis set of \citet{WoDu94} for all atoms. The ($3s3p2d1f$) set of bond functions defined by \citet{TaPa92} are added and placed at mid-distance between the c.m. of $\mathrm{CH{}^{+}}$ and $\mathrm{He}$. 

The basis set superposition errors (BSSE) was corrected at all geometries following the \citet{BoBe70} counterpoise procedure. Our PES has a global minimum of $537$\,cm$^{-1}$ at $R=4.05$\,Bohr and $\theta=84^\circ$. This value is consistent with that obtained by \citet{StVo08} with the BCCD(T) method and a aug-cc-pVQZ basis set, i.e., $513$\,cm$^{-1}$ at $R=4.1$\,Bohr and $\theta=86^\circ$. Indeed, our well depth is lower than their value.

To perform the dynamical calculations, the basic inputs required by the MOLSCAT package \citep{HuGr94} were obtained by expanding the interaction potential in terms of Legendre polynomials as
\begin{equation}
V(r=r_{\mathrm{e}},R,\theta)=\sum_{\lambda}V_{\lambda}(R)P_{\lambda}(cos\,\theta).\label{eq1}
\end{equation}
The calculated surface is well reproduced by the analytical potential over the entire grid of used coordinate points. The standard deviation between the analytical and the calculated surface remains below $1.0$\%. Further details concerning calculations related to the interaction potential can be found in the paper by \citet{HaOJ08}.

\section{Cross sections}

The quantum mechanical close coupling approach \citep{ArDa60} implemented in the MOLSCAT code was used to calculate state to state rotational integral cross sections for values of $J$ ranging from $0$ to $10$, and a total energy up to $10\,000\,\mathrm{cm^{-1}}$. The energy range was carefully spanned in order to account for resonances. 
The incremental steps were chosen as follows: From $0$ to $50\,\mathrm{cm^{-1}}$,
they were set to $0.1\,\mathrm{cm^{-1}}$, from $50\,\mathrm{cm^{-1}}$
to $100\,\mathrm{cm^{-1}}$ to $0.2\,\mathrm{cm^{-1}}$, from $100\,\mathrm{cm^{-1}}$
to $200\,\mathrm{cm^{-1}}$ to $0.5\,\mathrm{cm^{-1}}$, from $200\,\mathrm{cm^{-1}}$
to $1200\,\mathrm{cm^{-1}}$ to $1\,\mathrm{cm^{-1}}$, from $1200\,\mathrm{cm^{-1}}$
to $1400\,\mathrm{cm^{-1}}$ to $2\,\mathrm{cm^{-1}}$, from $1400\,\mathrm{cm^{-1}}$
to $1500\,\mathrm{cm^{-1}}$ to $5\,\mathrm{cm^{-1}}$, from $1500\,\mathrm{cm^{-1}}$
to $2000\,\mathrm{cm^{-1}}$ to $10\,\mathrm{cm^{-1}}$, from $2000\,\mathrm{cm^{-1}}$
to $2500\,\mathrm{cm^{-1}}$ to $25\,\mathrm{cm^{-1}}$, from $2500\,\mathrm{cm^{-1}}$
to $3000\,\mathrm{cm^{-1}}$ to $50\,\mathrm{cm^{-1}}$, from $3000\,\mathrm{cm^{-1}}$
to $5000\,\mathrm{cm^{-1}}$ to $100\,\mathrm{cm^{-1}}$, and finally
from $5000\,\mathrm{cm^{-1}}$ to $10\,000\,\mathrm{cm^{-1}}$ to $200\,\mathrm{cm^{-1}}$. 
To include all open channels and some closed channels, we have set $J_{max}=15$ for $E\leq1000\,\mathrm{cm^{-1}}$, and $J_{max}=17$ for $E>1000\,\mathrm{cm^{-1}}$. This corresponds to a rotational basis set of adequate size for a good accuracy in the calculated cross sections. The input parameters required by MOLSCAT are displayed in Table \ref{T1}. These parameters were fixed after we have performed some tests to ensure the convergence of cross sections for energies up to $10\,000\,\mathrm{cm^{-1}}$. The coupled equations were conveniently solved using the propagator of \citet{Man86}.

\begin{figure}
\caption{Deexcitation cross sections of $\mathrm{CH^{+}}$ by collision with $\mathrm{He}$ as a function of the kinetic energy.}
\resizebox{\hsize}{!}{\includegraphics{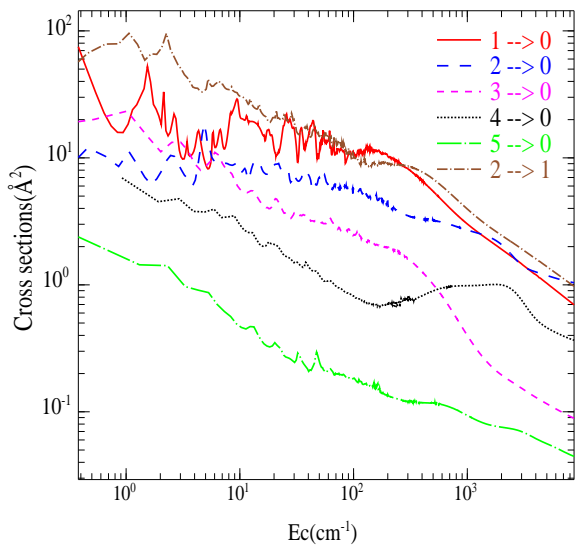}}
\label{F1}
\end{figure}

\begin{figure}
\caption{Rotational quenching cross sections of $\mathrm{CH^{+}}$ by collision with $\mathrm{He}$ as a function of the kinetic energy.}
\resizebox{\hsize}{!}{\includegraphics{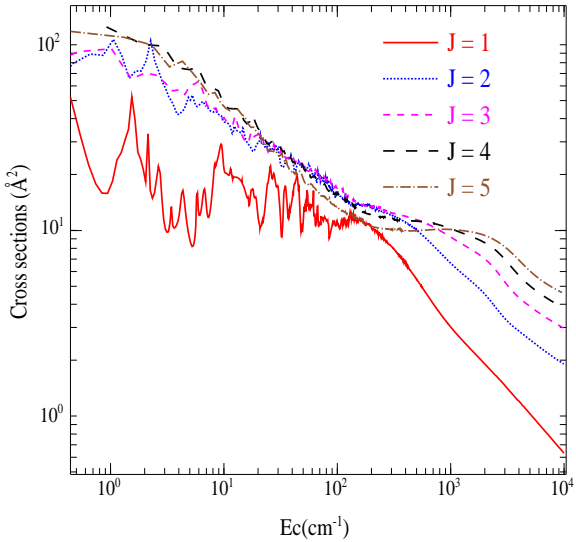}}
\label{F2}
\end{figure}

Figure \ref{F1} presents the energy variation of the $\mathrm{CH^{+}}-\mathrm{He}$ collisional deexcitation cross sections for the transitions $J\rightarrow 0$, for $J=1-5$ and the transition $2\rightarrow 1$. Up to almost $600\,\mathrm{cm^{-1}}$, resonances can be seen in the cross sections. These are due to the global minimum located at $4.05\,\mathrm{Bohr}$ and whose well depth is $\sim537\,\mathrm{cm^{-1}}$. As one can also see from the figure, the cross sections for the transition $1\rightarrow 0$ are larger than those with $J>1\rightarrow 0$ almost over the entire range of energies. Some exceptions occur with the transition $2\rightarrow 0$ for which cross sections are slightly larger around $5$\,cm$^{-1}$ and for $E_C \ga 1000$\,cm$^{-1}$. It is also obvious from the figure that the cross sections for the transition $2\rightarrow 1$ are larger than those of the other transitions plotted but remain of similar magnitude as those of the transition $1 \rightarrow 0$ with increasing energy. Turning our attention to Fig.~\ref{F2}, it can be seen that our total cross sections are consistent both in shape and magnitude with the quenching cross sections calculated by \citet{StVo08} for energies lower than $1000$\,cm$^{-1}$. 

An analysis of the excitation cross sections has been carried out by \citet{HaOJ08}. It was pointed out in that for energies lower than $600$\,cm$^{-1}$, the cross sections for the transition $0\rightarrow1$ are larger than those for $0\rightarrow2$ while for energies greater than that, the converse holds. Similar observations were also made for the transition $1\rightarrow2$ and $0\rightarrow2$ for an energy of $\sim150\,\mathrm{cm^{-1}}$. Indeed, for energies greater than that value, the transition $1\rightarrow2$ is less favored \citep[see also][]{HaOJ08}. It should be noticed that there is a propensity towards $\Delta J$ even parity transitions for almost the entire range of energy.

\begin{table*}
\caption{\label{T1}MOLSCAT parameters used in the present calculations.
Be and De are from \citet{HuHe79}.}
\begin{center}\begin{tabular}{cccc}
\hline 
INTFLG=6&
STEPS=30&
OTOL=0.01&
DTOL=0.1\tabularnewline
\hline
Be=14.1776\,cm$^{-1}$&
De=0.0014\,cm$^{-1}$&
JMAX=17&
RMIN=3\,Bohr, RMAX=50\,Bohr\tabularnewline
\hline
\end{tabular}\end{center}
\end{table*}

\section{Collisional rates}

\begin{figure}
\caption{Calculated downward rate coefficients for selected
transitions as a function of the kinetic temperature.}
\resizebox{\hsize}{!}{\includegraphics{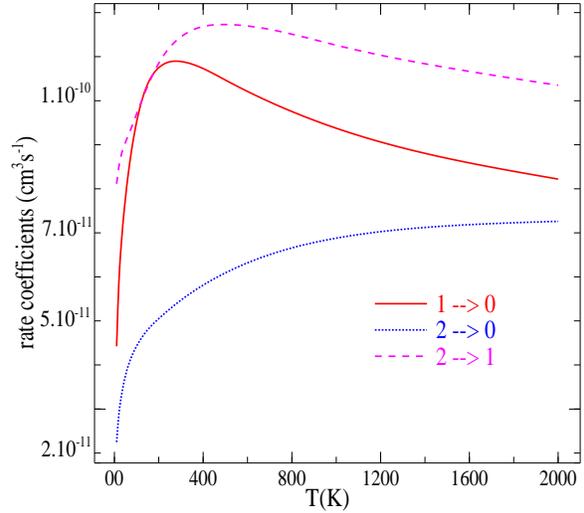}}
\label{F3}
\end{figure}

Downward rate coefficients are obtained by averaging the cross sections $\sigma_{J\rightarrow J'}$ over a Maxwell-Boltzmann distribution of kinetic energies,
\begin{equation}
q_{J\rightarrow J'}(T)=\left(\frac{8\beta^{3}}{\pi\mu}\right)\int_{0}^{\infty}E_{C}\sigma_{J\rightarrow J'}(E)e^{-\beta E_{C}}dE_{C}\label{eq2} \ ,
\end{equation}
where $T$ is the kinetic temperature, $\mu=3.06139\,\mathrm{a.u.}$ is the reduced mass of the $\mathrm{CH{}^{+}-He}$ collision partners, $\beta=\frac{1}{k_{B}T}$ ($k_{B}$ is the Boltzmann constant) and $E_{C}=E-E_{J}$ is the relative kinetic energy. Table \ref{T2} displays the results at selected temperatures. Additional numbers may be obtained upon request to the authors. Figure \ref{F3} illustrates the variation of $q_{J\rightarrow J'}(T)$ with the kinetic temperature. The general trends observed earlier by \citet{HaOJ08} for these numbers at low temperature remain. As one can see from the figure, the magnitude of the rate coefficients for the transition $2\rightarrow1$ dominates followed by the $1\rightarrow0$ which is dominant over $2\rightarrow0$.

\begin{table*}
\caption{\label{T2}Downward rate coefficients of rotational levels
of CH$^{+}$ in collision with He as a function of kinetic temperature
(in units of cm$^{3}$\,s$^{-1}$).}
\begin{center}\begin{tabular}{ccccccc}
\hline 
Initial level&
Final level&
\multicolumn{5}{c}{Rate Coefficients}\tabularnewline
\hline
J&
J'&
300\,K&
500\,K&
1000\,K&
1500\,K&
2000\,K\tabularnewline
 1 &
 0 &
1.0892(-10)&
1.0474(-10)&
9.3755(-11)&
8.6919(-11)&
8.2199(-11)\tabularnewline
 2 &
 0 &
5.4712(-11)&
6.0895(-11)&
6.8789(-11)&
7.1572(-11)&
7.2620(-11)\tabularnewline
 2 &
 1 &
1.1448(-10)&
1.1728(-10)&
1.1262(-10)&
1.0753(-10)&
1.0347(-10)\tabularnewline
 3 &
 0 &
2.1373(-11)&
1.9118(-11)&
1.4804(-11)&
1.2486(-11)&
1.1121(-11)\tabularnewline
 3 &
 1 &
8.8407(-11)&
1.0579(-10)&
1.3046(-10)&
1.4071(-10)&
1.4494(-10)\tabularnewline
 3 &
 2 &
7.4955(-11)&
8.2783(-11)&
8.9974(-11)&
9.1966(-11)&
9.2355(-11)\tabularnewline
 4 &
 0 &
1.2349(-11)&
1.6741(-11)&
2.4127(-11)&
2.7741(-11)&
2.9365(-11)\tabularnewline
 4 &
 1 &
1.8445(-11)&
1.8632(-11)&
1.7816(-11)&
1.7104(-11)&
1.6602(-11)\tabularnewline
 4 &
 2 &
1.0329(-10)&
1.2071(-10)&
1.4877(-10)&
1.6297(-10)&
1.7013(-10)\tabularnewline
 4 &
 3 &
4.4272(-11)&
5.3386(-11)&
6.6319(-11)&
7.2883(-11)&
7.6588(-11)\tabularnewline
 5 &
 0 &
1.8893(-12)&
2.1234(-12)&
2.4700(-12)&
2.6880(-12)&
2.8346(-12)\tabularnewline
 5 &
 1 &
2.6285(-11)&
3.2428(-11)&
4.5476(-11)&
5.3674(-11)&
5.8207(-11)\tabularnewline
 5 &
 2 &
9.2178(-12)&
1.1060(-11)&
1.3795(-11)&
1.5314(-11)&
1.6233(-11)\tabularnewline
 5 &
 3 &
9.5934(-11)&
1.1444(-10)&
1.4622(-10)&
1.6408(-10)&
1.7437(-10)\tabularnewline
 5 &
 4 &
2.6196(-11)&
3.4002(-11)&
4.8021(-11)&
5.6663(-11)&
6.2226(-11)\tabularnewline
 6 &
 0 &
4.5865(-12)&
5.5789(-12)&
8.2688(-12)&
1.0468(-11)&
1.1906(-11)\tabularnewline
 6 &
 1 &
1.7118(-12)&
2.3213(-12)&
3.6910(-12)&
4.7426(-12)&
5.4934(-12)\tabularnewline
 6 &
 2 &
2.7322(-11)&
3.4211(-11)&
4.9685(-11)&
6.0782(-11)&
6.7859(-11)\tabularnewline
 6 &
 3 &
4.5014(-12)&
6.5071(-12)&
1.0643(-11)&
1.3429(-11)&
1.5279(-11)\tabularnewline
 6 &
 4 &
8.6611(-11)&
1.0467(-10)&
1.3755(-10)&
1.5779(-10)&
1.7065(-10)\tabularnewline
 6 &
 5 &
1.7971(-11)&
2.4058(-11)&
3.6692(-11)&
4.5470(-11)&
5.1550(-11)\tabularnewline
 7 &
 0 &
2.9225(-13)&
4.2016(-13)&
7.6333(-13)&
1.0737(-12)&
1.3164(-12)\tabularnewline
 7 &
 1 &
8.1105(-12)&
1.0045(-11)&
1.5631(-11)&
2.0767(-11)&
2.4497(-11)\tabularnewline
 7 &
 2 &
1.5290(-12)&
2.2415(-12)&
4.1090(-12)&
5.7047(-12)&
6.9146(-12)\tabularnewline
 7 &
 3 &
2.4418(-11)&
3.1705(-11)&
4.8123(-11)&
6.0615(-11)&
6.9304(-11)\tabularnewline
 7 &
 4 &
4.0769(-12)&
5.6780(-12)&
9.7084(-12)&
1.2814(-11)&
1.5021(-11)\tabularnewline
 7 &
 5 &
7.3617(-11)&
9.2301(-11)&
1.2673(-10)&
1.4868(-10)&
1.6326(-10)\tabularnewline
 7 &
 6 &
1.5508(-11)&
2.0148(-11)&
3.0772(-11)&
3.8740(-11)&
4.4530(-11)\tabularnewline
 8 &
 0 &
1.4359(-12)&
1.8300(-12)&
3.0413(-12)&
4.3682(-12)&
5.4488(-12)\tabularnewline
 8 &
 1 &
9.0915(-13)&
1.2019(-12)&
1.9538(-12)&
2.6903(-12)&
3.3078(-12)\tabularnewline
 8 &
 2 &
7.7361(-12)&
1.0389(-11)&
1.7651(-11)&
2.4520(-11)&
2.9880(-11)\tabularnewline
 8 &
 3 &
2.3727(-12)&
3.1083(-12)&
5.0438(-12)&
6.8186(-12)&
8.2340(-12)\tabularnewline
 8 &
 4 &
2.0407(-11)&
2.8012(-11)&
4.5156(-11)&
5.8453(-11)&
6.8144(-11)\tabularnewline
 8 &
 5 &
5.3564(-12)&
6.7877(-12)&
1.0395(-11)&
1.3359(-11)&
1.5551(-11)\tabularnewline
 8 &
 6 &
5.9379(-11)&
7.8105(-11)&
1.1414(-10)&
1.3805(-10)&
1.5436(-10)\tabularnewline
 8 &
 7 &
1.6382(-11)&
1.9895(-11)&
2.8486(-11)&
3.5245(-11)&
4.0284(-11)\tabularnewline
 9 &
 0 &
3.1749(-13)&
4.1519(-13)&
6.0967(-13)&
7.9911(-13)&
9.6908(-13)\tabularnewline
 9 &
 1 &
2.2637(-12)&
3.2040(-12)&
6.0848(-12)&
9.2677(-12)&
1.2005(-11)\tabularnewline
 9 &
 2 &
1.6464(-12)&
2.1416(-12)&
3.1582(-12)&
4.1187(-12)&
4.9529(-12)\tabularnewline
 9 &
 3 &
6.3038(-12)&
9.4243(-12)&
1.7913(-11)&
2.5792(-11)&
3.2072(-11)\tabularnewline
 9 &
 4 &
3.5314(-12)&
4.4470(-12)&
6.3778(-12)&
8.1075(-12)&
9.5331(-12)\tabularnewline
 9 &
 5 &
1.5792(-11)&
2.3335(-11)&
4.0997(-11)&
5.4987(-11)&
6.5343(-11)\tabularnewline
 9 &
 6 &
6.9032(-12)&
8.4680(-12)&
1.1802(-11)&
1.4466(-11)&
1.6473(-11)\tabularnewline
 9 &
 7 &
4.6271(-11)&
6.3763(-11)&
9.9734(-11)&
1.2494(-10)&
1.4252(-10)\tabularnewline
 9 &
 8 &
1.8795(-11)&
2.1639(-11)&
2.8624(-11)&
3.4250(-11)&
3.8524(-11)\tabularnewline
10 &
 0 &
3.7206(-13)&
5.7888(-13)&
1.2592(-12)&
2.0545(-12)&
2.7827(-12)\tabularnewline
10 &
 1 &
8.7732(-13)&
1.2033(-12)&
1.7474(-12)&
2.2194(-12)&
2.6452(-12)\tabularnewline
10 &
 2 &
1.9744(-12)&
3.2175(-12)&
7.1936(-12)&
1.1441(-11)&
1.5137(-11)\tabularnewline
10 &
 3 &
2.2421(-12)&
3.0153(-12)&
4.3132(-12)&
5.3982(-12)&
6.3360(-12)\tabularnewline
10 &
 4 &
4.7195(-12)&
7.8986(-12)&
1.7002(-11)&
2.5507(-11)&
3.2383(-11)\tabularnewline
10 &
 5 &
4.3451(-12)&
5.6292(-12)&
7.7712(-12)&
9.4441(-12)&
1.0795(-11)\tabularnewline
10 &
 6 &
1.1332(-11)&
1.8149(-11)&
3.5276(-11)&
4.9542(-11)&
6.0390(-11)\tabularnewline
10 &
 7 &
8.1395(-12)&
1.0071(-11)&
1.3538(-11)&
1.6073(-11)&
1.7923(-11)\tabularnewline
10 &
 8 &
3.5458(-11)&
5.1093(-11)&
8.5561(-11)&
1.1114(-10)&
1.2939(-10)\tabularnewline
10 &
 9 &
2.1861(-11)&
2.4448(-11)&
3.0405(-11)&
3.5149(-11)&
3.8707(-11)\tabularnewline
\hline
\end{tabular}\end{center}
\end{table*}

\section{Summary}

Using a previously computed PES \citep{HaOJ08}, we have obtained results of a quantum mechanical close coupling calculation of integral cross sections for transitions between the lower rotational levels of $\mathrm{CH{}^{+}}$ induced by collisions with $\mathrm{He}$. The cross sections were averaged over a Maxwell-Boltzmann distribution of kinetic energies to determine downward rate coefficients. The kinetic temperature spans a wide range of values up to $2000\,\mathrm{K}$. The spectroscopic parameters obtained at high temperature exhibit the general trends for such quantities. The rate coefficients will help to understand and interpret astrophysical observations in connection with the Herschel Space Observatory mission.


\begin{acknowledgements}
Author LCOO acknowledges with thanks the financial support
of the Abdus Salam International Centre for Theoretical Physics, Office
of External Activities (ICTP-OEA ) under NET45 Programme.
\end{acknowledgements}

\bibliographystyle{aa}
\bibliography{pst}

\end{document}